\documentclass[
 reprint,
 onecolumn,
superscriptaddress,
 amsmath,amssymb,
 aps,
 pra,
 nolongbibliography,
nofootinbib
]{revtex4-2}
\newcommand{\sro}[0]{{\sqrt{\rho}}}
\usepackage[dvipsnames]{xcolor}

\usepackage{graphicx}
\usepackage{dcolumn}
\usepackage{soul}
\usepackage{bm}
\usepackage{hyperref}
\usepackage{comment}
\usepackage{lipsum}
\usepackage{soul}
\begin{document}

\title{Propagation properties and stability of dark solitons\\in weakly interacting Bose-Bose droplets}
\author{Jakub Kopyci\'{n}ski}
\email{jkopycinski@cft.edu.pl}
\affiliation{Center for Theoretical Physics, Polish Academy of Sciences, Al. Lotnik\'{o}w 32/46, 02-668 Warsaw, Poland}
\author{Bu\u{g}ra T\"uzemen}
\affiliation{Center for Theoretical Physics, Polish Academy of Sciences, Al. Lotnik\'{o}w 32/46, 02-668 Warsaw, Poland}
\author{Wojciech G\'{o}recki}
\affiliation{Faculty of Physics, University of Warsaw, Pasteura 5, 02-093 Warsaw, Poland}
\author{Krzysztof Paw{\l}owski}
\affiliation{Center for Theoretical Physics, Polish Academy of Sciences, Al. Lotnik\'{o}w 32/46, 02-668 Warsaw, Poland}
\author{Maciej \L ebek}
\affiliation{Center for Theoretical Physics, Polish Academy of Sciences, Al. Lotnik\'{o}w 32/46, 02-668 Warsaw, Poland}
\affiliation{Faculty of Physics, University of Warsaw, Pasteura 5, 02-093 Warsaw, Poland}
\date{\today}

\begin{abstract}
We investigate dark solitons in two-component Bose systems with competing interactions in one dimension. Such a system hosts a liquid phase stabilized by the beyond-mean field corrections. Using the generalized Gross-Pitaevskii equation, we reveal the presence of two families of solitonic solutions. The solitons in both of them can be engineered to be arbitrarily wide. One family of solutions, however, has got an anomalous dispersion relation and our analyses show one of its branches is unstable. We find the presence of a critical velocity demarcating the stable from unstable solutions. 
Nonetheless, grey anomalous solitons are able to exist inside quantum droplets and can be treated as solitonic excitations thereof. 
\end{abstract}

\maketitle

\begin{figure}[h!]
    \centering
    \includegraphics[trim=15 10 0 0,clip]{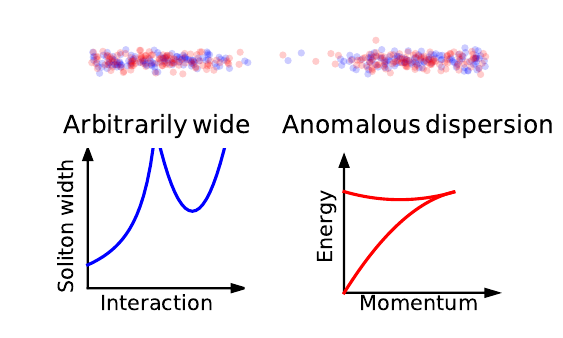}
    \caption{Graphical abstract: we look into dark solitons in two-component Bose gases
    (species coloured red and blue) with a beyond mean-field approach. Our studies show that dark solitons can be arbitrarily wide in such systems. Moreover, we find a class of solutions with an anomalous dispersion relation and perform a stability analysis of these soloutons. We predict the moving solitons can exist in the quantum droplets as their excitations.}
    \label{fig:abstract}
\end{figure}

\section{Introduction}

Ultracold atomic systems with competing interactions have been a subject of an extensive theoretical research, powered up by various experiments. The addition of corrections for quantum fluctuations~\cite{Petrov_2015} to the mean-field theory led to the correct description of the emergence of self-bound objects called quantum droplets in two-component systems~\cite{Cabrera_2018, Cheiney_2018, Semeghini_2018, Ferioli_2019, DErrico_2019}.
A quite unexpected appearance of droplets in dipolar gases~\cite{Kadau_2016, Schmitt_2016, Ferrier-Barbut_2016} also became explicable by this effect. 
All in all, the mean-field approach tumbled down in these two regimes and the use of its version with LHY corrections~\cite{Lee_1957a, Lee_1957b} -- known as the generalized Gross-Pitaevskii equation (GGP), became crucial.

The GGP~\cite{Petrov_2016} proves its usefulness in the weakly-interacting regime~\cite{Parisi_2019} of Bose-Bose mixtures. It has been employed to study the properties of quantum droplets~\cite{Astrakharchik_2018}, including elementary excitations~\cite{Tylutki_2020} like breathing modes~\cite{Parisi_2020}, even in the case of unequal number of bosons of each species~\cite{Flynn_2023a, Flynn_2023b}. 

Besides quantum droplets, bosonic mixtures may uncover also bright solitons~\cite{Liu_2022}, mixed bubbles~\cite{Sturmer_2022} near the misciblie-immiscible threshold, experimentally seen dark-bright solitons~\cite{Yan_2015, Keverekidis_2016}, predicted also for highly magnetic species~\cite{Adhikari_2014}.

There is a variety of already known dark-soliton-like solutions of the GGP in the miscible regime, such as kink-type solitons~\cite{Shukla_2021}, dark quantum droplets~\cite{Edmonds_2023} and standard dark solitons~\cite{Katsimiga_2023a}. 

Let us briefly describe the most important properties of these excitations. The kink-type soliton density profile has got two different asymptotic values with a rapidly varying region located at its origin. In this case, the phase of the wave function remains constant. 

Dark quantum droplets, on the other hand, have a shape of an inverted quantum droplet. Their phase pattern is non-trivial, with $-\pi/2$ phase on one side of the depletion, $\pi/2$ on the other one with an intermediate step in the wide depleted density region of phase equal to $0$. All three regions are smoothly connected.

Standard dark solitons can be divided into fully and partially depleted -- black and grey correspondingly. Black solitons have a $\pi$-jump in their phase. The phase of grey solitons changes smoothly and the total phase difference is smaller than $\pi$. All in all, they share the features of dark solitons in single-component Bose gases~\cite{Jackson_1998, Sato_2016}. 

Yet, there is still another type of solitons, so far  found only in a beyond-LHY description called LLGPE of 1D Bose-Bose mixtures~\cite{Kopycinski_2023b} and dipolar Bose gases~\cite{Kopycinski_2023a}, but had been retrieved as the solutions of some differential equations~\cite{Berestycki_1983} and called ``solitonlike bubbles" there. These anomalous solitons can be arbitrarily wide, are never fully depleted and have a constant phase profile. Moreover, they have a peculiar dispersion relation with an additional subbranch and a cusp. As they were not found in~\cite{Edmonds_2023}, one may think they appear due to beyond-LHY contributions to the method.

In this article, we derive the solitonic solution of the GGP and reveal the presence of anomalous solitons. We also check the stability of the solutions and show that dark solitons can exist inside quantum droplets~(see figure~\ref{fig:abstract}).

\section{Framework} We look into a weakly interacting two-component gas of $\tilde{N}$ bosons in 1D. The intraspecies interaction is repulsive, whereas the intercomponent ones are attractive. Their masses are the same, i.e. $m_\uparrow=m_\downarrow=m$ with $\sigma=\{\uparrow,\downarrow\}$ denoting the component. The energy density functional for this system is given by~\cite{Petrov_2016}:
\begin{equation}
    \mathcal{E}_{\rm int}[\tilde\rho_\uparrow, \tilde\rho_\downarrow]=\frac{\left(\sqrt{g_{\uparrow\uparrow}}\tilde\rho_\uparrow-\sqrt{g_{\downarrow\downarrow}}\tilde\rho_\uparrow\right)^2}{2}+\left(g_{\uparrow\downarrow}\sqrt{g_{\uparrow\uparrow}g_{\downarrow\downarrow}}+g_{\uparrow\uparrow}g_{\downarrow\downarrow}\right)\frac{\left(\sqrt{g_{\uparrow\uparrow}}\tilde\rho_\uparrow+\sqrt{g_{\downarrow\downarrow}}\tilde\rho_\uparrow\right)^2}{\left(g_{\uparrow\uparrow}+g_{\downarrow\downarrow}\right)^2}
    -\frac{2\sqrt{m}}{3\pi\hbar}\left(g_{\uparrow\uparrow}\tilde\rho_\downarrow+g_{\downarrow\downarrow}\tilde\rho_\downarrow\right)^{3/2},
\end{equation}
where $\tilde\rho_\sigma$ is the $\sigma$-component density and $g_{\sigma\sigma'}$ for $\sigma=\sigma'$ is the intracomponent interaction strength and the intercomponent interaction strength otherwise, i.e. when $\sigma\neq\sigma'$. 

In the miscible system without any external confinement the densities are tied up via the condition $\tilde\rho_\uparrow/\tilde\rho_\downarrow=\sqrt{g_{\downarrow\downarrow}/g_{\uparrow\uparrow}}$~\cite{Petrov_2015}. 
This leads to a simplification of the energy density functional $\mathcal{E}_{\rm int}$. An approach based on the local density approximation enables us to use $\mathcal{E}_{\rm int}$ and write the generalized Gross-Pitaevskii equation for the wave function $\Phi(\tilde{x}, \tilde{t})$ as follows~\cite{Petrov_2016}:

\begin{equation}
    i\hbar\partial_{\tilde t}\tilde \Phi=-\frac{\hbar^2}{2m}\partial^2_{\tilde x}\tilde\Phi+\frac{2 \sqrt{g_{\uparrow\uparrow} g_{\downarrow\downarrow}} \delta g}{(\sqrt{g_{\uparrow\uparrow} }+\sqrt{g_{\downarrow\downarrow}})^2}|\tilde \Phi|^2\tilde \Phi
    -\frac{\sqrt{m}}{\pi\hbar}(g_{\uparrow\uparrow}g_{\downarrow\downarrow})^{3/4}|\tilde \Phi|\tilde \Phi,
    \label{eq:ggp}
\end{equation}
where $\tilde{\Phi}(\tilde{x},\tilde{t})$ is related to single-component wave functions $\tilde{\Phi}_\sigma$ via $\tilde{\Phi}_\sigma(\tilde{x},\tilde{t})=g_{\bar{\sigma}\bar{\sigma}}^{1/4} \tilde{\Phi}(\tilde{x},\tilde{t})\sqrt{\sqrt{g_{\uparrow\uparrow}}+\sqrt{g_{\downarrow\downarrow}}}$, where $\bar{\uparrow}=\downarrow$ and $\bar{\downarrow}=\uparrow$. Moreover, we have defined $\delta g=g_{\uparrow\downarrow}+\sqrt{g_{\uparrow\uparrow} g_{\downarrow\downarrow}}$, which is assumed to be positive throughout the work.

We now introduce the units of length $x_0 \equiv \frac{\pi \hbar^2}{m}\frac{\sqrt{2 \delta g}}{\sqrt{g_{\uparrow\uparrow} g_{\downarrow\downarrow}}(\sqrt{g_{\uparrow\uparrow} }+\sqrt{g_{\downarrow\downarrow}})}$ , time $t_0=\hbar/m x_0^2$, energy $E_0=\hbar/t_0$ and normalization factor of the wave function $ \Phi_0 = \frac{(\sqrt{g_{\uparrow\uparrow} }+\sqrt{g_{\downarrow\downarrow}})^{3/2}}{\sqrt{\pi x_0}(2\delta g)^{3/4}}$~\cite{Tylutki_2020}.
With $ \tilde t= t_0 t$, $\tilde x= x_0 x$, $ \tilde\Phi=\Phi_0 \Phi$ and $\tilde E=E_0 E$, we can rewrite~\eqref{eq:ggp} in the dimensionless form~\cite{Astrakharchik_2018}:
\begin{equation}
       i\partial_t\Phi(x,t)=-\frac{1}{2}\partial^2_x\Phi(x,t)+|\Phi(x,t)|^2\Phi(x,t)
       -|\Phi(x,t)|\Phi(x,t).
       \label{eq:dimless}
\end{equation}
The normalization condition of the wave function $\Phi$ is related to the real number of atom in the system $\tilde{N}$ via $N=\int_{-L/2}^{L/2}|\Phi|^2dx=\tilde{N}\pi(2\delta g)^{3/2}/(\sqrt{g_{\uparrow\uparrow}}+\sqrt{g_{\downarrow\downarrow}})^3$~\cite{Tylutki_2020}, assuming a box of size $L$ with periodic boundary conditions.

\section{Results}
\subsection{Speed of sound and equilibrium density}
The stable and unstable liquid phases in Bose-Bose mixtures are demarcated by the point where the speed of sound becomes imaginary~\cite{Parisi_2019}. We use the following time-dependent Ansatz to linearize~\eqref{eq:dimless}:
\begin{equation}
    \Phi(x,t)=\left[\Phi_0+\delta\Phi(x,t)\right]e^{-i\mu_0 t},
\end{equation}
where $\Phi_0=\sqrt{N/L}$, $\mu_0=\frac{N}{L}-\sqrt{\frac{N}{L}}$ is the chemical potential corresponding to the constant density profile and $\delta\Phi$ is a small perturbation.
Then, with $\omega_\rho=\rho_0-\frac{1}{2}\sqrt{\rho_0}$, we obtain a set of Bogoliubov-de Gennes (BdG) equations
\begin{equation}
     i\partial_t \begin{pmatrix}\delta\Phi\\\delta\Phi^*
    \end{pmatrix}
    =
    \begin{pmatrix}-\partial_x^2/2+\omega_\rho &\omega_\rho\\-\omega_\rho&-\left(-\partial_x^2/2+\omega_\rho\right)
    \end{pmatrix}\begin{pmatrix}\delta\Phi\\\delta\Phi^*
    \end{pmatrix}.
\end{equation}
Assuming that $\delta\Phi=ue^{(ikx-i\omega t)}+v^*e^{-(ikx-i\omega t)}$~\cite{Pitaevskii_2016}, we diagonalize the Hamiltonian matrix an obtain the following eigenfrequencies 
\begin{equation}
    \omega=\sqrt{\frac{k^4}{4}+k^2\left(\rho_0-\frac{1}{2}\sqrt{\rho_0}\right)}.
    \label{eq:eigenen}
\end{equation}
The spectrum is linear in the limit of low momenta, well approximated by the phonon dispersion law $\omega=ck$, where
\begin{equation}
    c = \sqrt{\rho_0-\frac{1}{2}\sqrt{\rho_0}}
    \label{eq:sound}
\end{equation}
is the speed of sound. On the other hand, when $k\gg1$,~\eqref{eq:eigenen} reduces to the free particle spectrum $\omega=k^2/2$.

From~\eqref{eq:sound}, we see the speed of sound is imaginary when $\rho_0<1/4$. It indicates that the stable-to-unstable liquid transition happens when $\rho_0=1/4\equiv\rho_\mathrm{ins}$ and this is a phonon instability.

The 
liquid phase is characterized by the presence of a minimum in the energy per particle function~\cite{Astrakharchik_2018, Parisi_2019}
\begin{equation}
    E/N=\frac{1}{2}\rho_0-\frac{2}{3}\rho_0^{1/2}.
\end{equation}

This minimum occurs when $\rho_0=4/9$ and we will further refer to this value as the equilibrium density $\rho_\mathrm{eq}$. When 
$N/L<\rho_\mathrm{eq}$, the system prefers to form a quantum droplet with a bulk density equal to $\rho_\mathrm{eq}$. This interpretation is valid if $N\gg1$. In such a case, the surface energy needed to form the droplet density profile is small in comparison to the bulk energy and therefore negligible.

\subsection{Solitonic solutions}
We now look for a solution of a dispersionless wave moving with velocity $v$ through an infinite constant background density. Such an object is represented by the following wave function:
\begin{subequations}
\begin{equation}
\Phi(x,t)\equiv \psi(x-vt)\exp\left(-i\mu t\right),
\label{eq:ansatz}
\end{equation}
where
\begin{equation}
\psi(\zeta)=\sqrt{\rho(\zeta)}\exp[i\phi(\zeta)]   
\end{equation}
\end{subequations}
and $\zeta$ is the comoving coordinate $\zeta=x-vt$.

One can insert~\eqref{eq:ansatz} to the GGP equation~\eqref{eq:dimless}. Then, we split a single complex non-linear differential equation into two -- the real part
\begin{subequations}
    \begin{equation}
    \mu\sro-v\phi'\sro+\frac{1}{2}(\sro)''-\frac{1}{2}(\phi')^2\sro-\sro^3+\sro^2=0
    \label{eq:real_pt}
\end{equation}
and the imaginary one:
\begin{equation}
  \phi''\sro^2+2\phi'(\sro)'\sro-2v(\sro)'\sro=0,
  \label{eq:imag_pt}
\end{equation}
\end{subequations}
where $(\cdot)':=\frac{d}{d\zeta}(\cdot)$.

\begin{figure}[t!]
    \centering
    \includegraphics{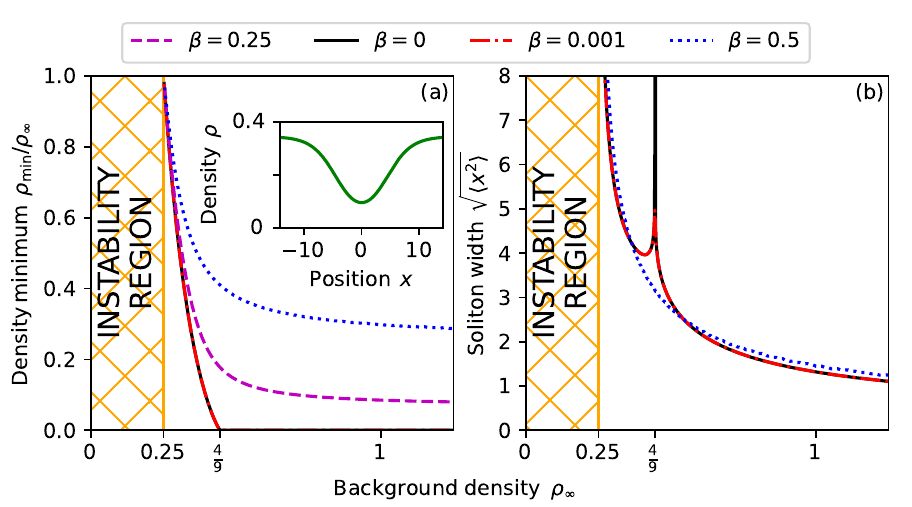}
    \caption{Dark soliton (a) density minima and (b) RMS widths as functions of the background density $\rho_\infty$ for different relative velocities $\beta=\{0,0.001, 0.25,0.5\}$. Inset in (a): density profile of an anomalous soliton ($\beta=0$ and $\rho_\infty=0.345$).}
    \label{fig:dep_and_w}
\end{figure}

Assuming that the density and phase are constant far away from the soliton, i.e. $\lim_{\zeta\to\pm\infty}\rho(\zeta)=\rho_{\infty}$ and $\lim_{\zeta\to\pm\infty}\phi(\zeta)=\pm\phi_{\infty}$ enables us to simplify the set of equations~\eqref{eq:real_pt} and~\eqref{eq:imag_pt} to:

\begin{subequations}
\begin{equation}
  \left(\frac{\rho'}{2}\right)^2+U(\rho)=0  
  \label{eq:de_rho}
\end{equation}
\begin{equation}
\phi'=v\left(1-\frac{\rho_\infty}{\rho}\right),
\label{eq:de_phi}
\end{equation}
\end{subequations}
with $U(\rho)$ given by:
\begin{equation}
    U(\rho)=(\rho-\rho_\infty)^2\left[v^2-\rho+\frac{2\rho(2\sqrt{\rho}+\sqrt{\rho_\infty})}{3(\sqrt{\rho}+\sqrt{\rho_\infty})^2}\right].
\end{equation}

We numerically\footnote{In principle, it is possible to proceed with the analytical calculations. Nevertheless, the solution is very complex and does not give much insight into the problem.} solve~\eqref{eq:de_rho} for a given $\rho_\infty$ and $v$ first and then use it to solve~\eqref{eq:de_phi}. 

Figure~\ref{fig:dep_and_w} shows the density minimum $\min \rho(x,t=0)$ and root mean square (RMS) width $\sqrt{\langle x^2\rangle}$ (assuming $\langle x\rangle=0$) of the solitonic excitations as functions of the background density. One can distinguish 3 regions there: (i) unstable liquid, (ii) anomalous, and (iii) the standard one.

In the unstable region, when $\rho_\infty<\rho_\mathrm{ins}$, even a small perturbation to a uniform density can cause a violent dynamics in the system. Thus, it is no surprise that there are no solitonic solutions there.

Otherwise, when $\rho_\mathrm{ins}<\rho_\infty<\rho_\mathrm{eq}$, we are in the anomalous regime. The inequality fulfils the necessary condition for the existence of an anomalous soliton~\cite{Berestycki_1983, Lin_2002}, namely $0<\rho_{\rm min}<\rho_\infty<\rho_{\rm eq}$ such that $U(\rho_{\min})=U(\rho_\infty)=0$ and $U(\rho)<0\,\forall\,\rho_{\rm min}\leqslant\rho\leqslant \rho_\infty$ as well as the stability of a uniform system, i.e. $c^2>0$. A motionless anomalous soliton is shown in the inset of figure~\ref{fig:dep_and_w}(a). Its characteristic feature is a partial depletion of density even if the soliton is motionless ($\beta\equiv v/c=0$). Another important property of the motionless soliton in the anomalous regime is a constant phase profile $\phi(\zeta;\beta=0)=const$.

\begin{figure}
    \centering
    \includegraphics{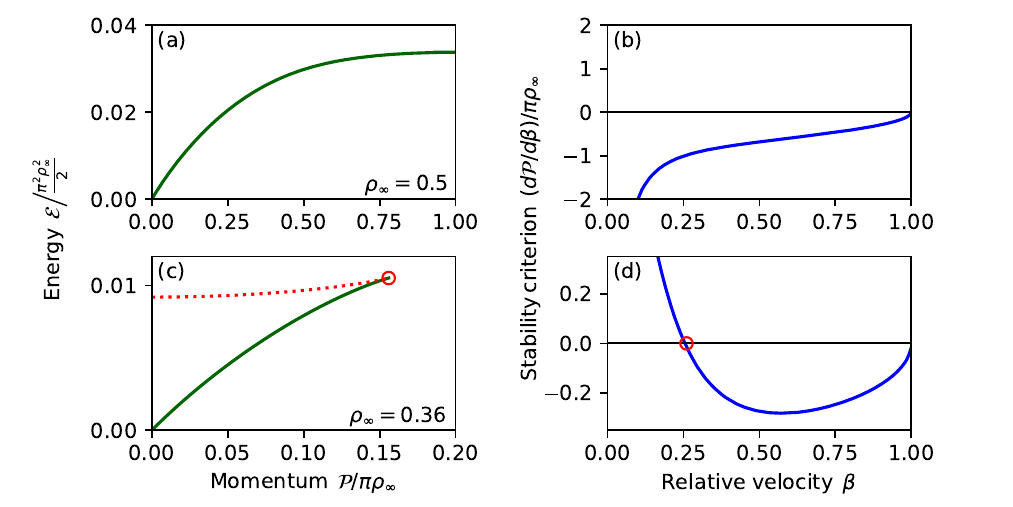}
    \caption{(a,c) Dispersion relation of solitons [standard -- top row, anomalous -- bottom row]. The red dashed line shows the upper subbranch. (b,d) Soliton stability criterion. Solitons are stable when $d\mathcal{P}/d\beta<0$. In the standard regime, the solitons are stable everywhere. In the anomalous one -- only above the critical velocity, which in this case of $\rho_\infty=0.36$ is numerically evaluated to be $\beta_{\rm cr}=0.2550(29)$. [Red circles in panels (c,d) mark the place, where the velocity of solitons equals to the critical velocity $\beta_{\rm cr}$.]}
    \label{fig:dispersion}
\end{figure}

In the last regime, $\rho_\infty>\rho_\mathrm{eq}$, the solitons have standard properties in terms of their density minimum and phase profile. Namely the density of a black ($\beta=0$) soliton reaches zero and there is a typical $\pi$-jump in the phase. 

As one can see in figure~\ref{fig:dep_and_w}(b), the motionless solitons -- both anomalous and standard ones -- become ultrawide when $\rho_\infty\to\rho_\mathrm{eq}$. It gives an opportunity to steer their size with the number of atoms in the system and interaction strengths $g_{\sigma\sigma'}$. The presence of arbitrarily wide solitons with a substantial density depletion in weakly interacting Bose-Bose mixtures is an opportunity for taking an \textit{in situ} absorption image of a dark soliton in such systems.

It is worth mentioning the kink-type solitons described in~\cite{Shukla_2021} occur exactly at $\rho_{\rm eq}$, i.e. $\lim_{\zeta\to\infty}\rho_{\rm kink}(\zeta)=\rho_{\rm eq}$ and $\lim_{\zeta\to-\infty}\rho_{\rm kink}(\zeta)=0$.

Another distinct property of the anomalous solitons, already described in~\cite{Kopycinski_2023a}, is the appearance of a subbranch in the dispersion relation. In figure~\ref{fig:dispersion}(a),~(c), we present the relation of the renormalized\footnote{We need this renormalization to properly evaluate the difference in energy between the homogeneous profile and the one with a soliton. Namely, we have to compute it in a finite box an only then go to the thermodynamic limit.} energy~\cite{Edmonds_2023}:
\begin{equation}
    \mathcal{E}=\int_{-\infty}^\infty\Bigg[\frac{1}{2}\left|\frac{d\psi}{d\zeta}\right|^2+\frac{1}{2}\left(\rho_\infty-\rho\right)^2
    -\frac{2}{3}\left(\rho^{3/2}-\frac{3}{2}\rho_\infty^{1/2}\rho+\frac{1}{2}\rho_\infty^{3/2}\right) \Bigg]d\zeta
\end{equation}
and regularized momentum~\cite{Kivshar_1998}:
\begin{equation}
    \mathcal{P}=v\int_{-\infty}^\infty\left(\rho-\rho_\infty\right)d\zeta-2\rho_\infty\phi_\infty.
\end{equation}

Also in this case, there is a violent change in the energy spectrum when we cross $\rho_\infty=4/9$ from above and enter the anomalous region. Namely, an additional subbranch altogether with a cusp appears.

It makes the effective mass $m_{\rm eff}= (d^2\mathcal{E}/d\mathcal{P}^2)^{-1}$ not properly defined due to the lack of derivative.

\begin{figure}[b]
    \centering
    \includegraphics{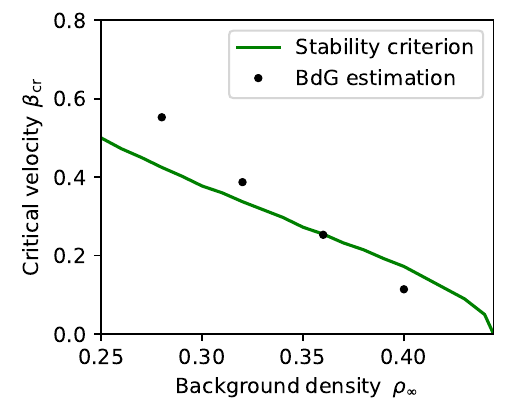}
    \caption{Critical velocity based on the stability criterion $d\mathcal{P}/d\beta<0$ (solid green line) and its estimation based on the Bogoliubov-de Gennes analysis of solitonic wave function.}
    \label{fig:stability}
\end{figure}

The stability analysis reveals another meaning of the two subbranches. The soliton is stable when $d\mathcal{P}/d\beta<0$~\cite{Barashenkov_1989}. In our case, this condition is fulfilled only for the lower branch of anomalous solitons. It means there is a critical velocity $\beta_{\rm cr}$ below which the anomalous soliton is unstable.

In figure~\ref{fig:dispersion}(d) we show the stability criterion and see the solitons are stable only above a certain velocity $\beta_{\rm cr}$, marked with a red circle. If we compare this picture with figure~\ref{fig:dispersion}(c), the upper subbranch corresponds to solitons moving with $ \beta\leqslant\beta_{\rm cr}$ (also marked with a red circles). Thus, the upper branch is the unstable one.

On the other hand, standard solitons with $\rho_\infty>\rho_{\rm eq}$, are always stable, which we can see in figure~\ref{fig:dispersion}(b), namely $\beta_{\rm cr}=0$. On the anomalous side, when we approach the standard regime ($\rho_\infty\to\rho_{\rm eq}^-$), we have $\beta_{\rm cr}\to0$, whereas while getting closer to the unstable one ($\rho_\infty\to\rho_{\rm ins}^+$), $\beta_{\rm cr}\to1/2$, which we show in figure~\ref{fig:stability}.

To give further evidence on the matter of the soliton stability, we numerically perform a Bogoliubov-de Gennes analysis of solitonic wave functions (see Appendix~\ref{app:B} for technical details). The lowest-state eigenenergy $\omega$ indicates whether or not the soliton is stable. We extract the critical velocity by finding the soliton velocity for which the BdG eigenvalue becomes real. The results are qualitatively consistent with the analysis based on the stability criterion from~\cite{Barashenkov_1989} (cf.\ figure~\ref{fig:stability}).

\section{Dark soliton in a quantum droplet}

\begin{figure}[h!]
    \centering
    \includegraphics{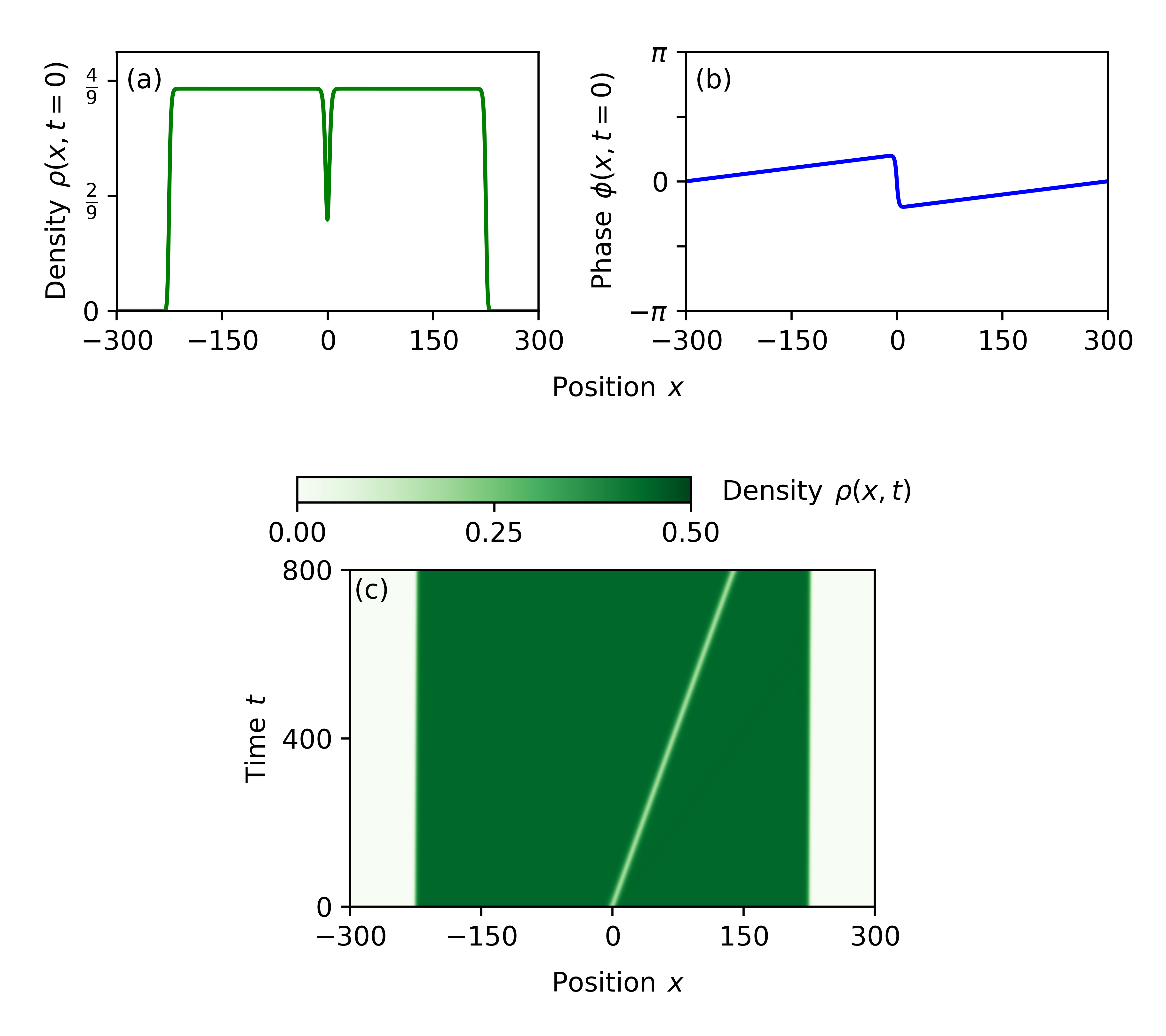}
    \caption{Dark soliton in a quantum droplet.
    Initial density (a) and phase (b) profiles as well as a space-time diagram of evolution (b) of a quantum droplet with a grey ($\beta=0.5$) soliton inside. The number of particles in the system $N=198.079$.}
    \label{fig:dropsol}
\end{figure}

The analysis of Bogoliubov excitations in quantum Bose-Bose droplets have revealed the presence of phonon modes~\cite{Tylutki_2020}.
We now check whether solitons
are present in quantum droplets as well. As the droplet is an inhomogeneous object, we use methods previously applied to trapped gases~\cite{Parker_2010}. We describe this approach below.

We take the following solution for a quantum droplet~\cite{Petrov_2016}
\begin{equation}
    \psi_{\rm QD}(x)=\frac{\sqrt{\rho_{\rm eq}}\mu/\mu_\mathrm{eq}}{1+\sqrt{1-\mu/\mu_{\rm eq}}\cosh\left(\sqrt{-2\mu} x\right)},
    \label{eq:psiQD}
\end{equation}
where $\mu_{\rm eq}=\rho_\mathrm{eq}-\sqrt{\rho_{\rm eq}}$ is the chemical potential corresponding to a homogeneous density profile with density $\rho=\rho_{\rm eq}$ and the number of particles in the droplet $N$ as a function of the chemical potential $\mu$ is given by~\cite{Tylutki_2020}:
\begin{equation}
N=\rho_{\rm eq}\sqrt{-\frac{2}{\mu_{\rm eq}}}\left[\ln\left(\frac{1+\sqrt{\mu/\mu_{\rm eq}}}{\sqrt{1-\mu/\mu_{\rm eq}}}\right)-\sqrt{\mu/\mu_{\rm eq}} \right].
\label{eq:droplet}
\end{equation}
Next, we modify the droplet wave function in order to get a dark soliton inside it.
We achieve it by simply multiplying~\eqref{eq:psiQD}
by the solitonic wave function $\psi(x)=\sqrt{\rho(x)}e^{i\phi(x)}$, where $\rho(x)$ and $\phi(x)$ are the solutions of Eqs.~\eqref{eq:de_rho} and~\eqref{eq:de_phi} for $\rho_\infty=\max |\psi_{\rm QD}(x)|^2$ and a given non-zero velocity $\beta\neq 0$. We will comment on the case $\beta=0$ later. Then, we normalize the overall wave function $\psi(x)\psi_{\rm QD}(x)$, impose periodic boundary conditions on the phase and  numerically evolve in real time (see Appendix~\ref{app:A} for more details).

Figure~\ref{fig:dropsol} shows us the results of these simulations. The initial density and phase profiles of a moving solitons are shown in figure~\ref{fig:dropsol}(a) and (b) correspondingly.

The space-time diagram in figure~\ref{fig:dropsol}(c)  show us how the anomalous motionless and moving solitons behave in quantum droplets. The dynamics is very stable.
We do not see any phonons or shock waves appearing.
The grey soliton in figure~\ref{fig:dropsol}(c)  indeed travels at $\beta=0.5$. When the droplet bulk is much larger than the soliton width, one should be able to use~\eqref{eq:droplet} along with the solution of Eqs.~\eqref{eq:de_rho} and~\eqref{eq:de_phi} to construct the wave function of a dark soliton-quantum droplet system.

Let us now return to the condition we made earlier about the non-zero soliton velocity. Are the motionless solitons not able to exist inside quantum droplets? According to our stability analysis, the motionless solitons are bound to be unstable.
We emphasize, however, that one still can achieve very wide stable solitons inside quantum droplets having a small, but finite velocity~$\beta$, as $\lim_{{\rho_\infty}\to{\rho_{\rm eq}^-}} {\beta_{\rm cr}}=0$.

\section{Conclusions}
We have shown the basic properties of a single dark soliton in a weakly interacting Bose-Bose mixture. There are two types of solitons: standard and anomalous ones. The standard motionless solitons are black and they have a $\pi$-jump in the phase, whereas the motionless do not have a fully depleted density nor a phase jump. In the anomalous regime, the solitons have also a peculiar dispersion relation with a cusp and subbranch. We found there is a critical velocity of the solitons $\beta_{\rm cr}$, below which the solitons are unstable. The stability analysis plays a crucial role in the possible experimental realization of solitons in quantum droplets. First, the unstable solitons will be suppressed in a potential experiment. Second, it is an opportunity to study the instability mechanism.

Our study disfavours the hypothesis that anomalous solitons are a beyond-LHY feature. 
Our observations~\cite{Kopycinski_2023a,Kopycinski_2023b} are consistent with the condition for the presence of anomalous solitons~\cite{Berestycki_1983, Lin_2002}, where
a preferred equilibrium density $\rho_\mathrm{eq}$ 
plays the key role in the existence of these objects. 

Most importantly, we show that this uncanny type of solitons can
exist inside a quantum droplet. According to our numerical simulations, 
the stable anomalous grey 
solitons might be experimentally observable in the droplets. 

A theoretical study of multiple-soliton systems is a natural way to extend our research, especially if it goes for two-soliton interactions.

\vspace{0.25cm}
\hypertarget{sec:data-avail}{\textit{Data availability ---}} All the numerical data necessary to reproduce figures, BdG data, and the results of simulations with the MUDGE toolkit (\url{https://gitlab.com/jakkop/mudge/-/releases/v03Apr2023}) are available under \url{https://doi.org/10.5281/zenodo.8427527}.

\begin{acknowledgments}
J.K., B.T., K.P., and M.Ł.\ acknowledge support from the (Polish) National
Science Center Grant No.~2019/34/E/ST2/00289.

We acknowledge the assistance from Center for Theoretical Physics of the Polish
Academy of Sciences which is a member of the National Laboratory of Atomic, Molecular and Optical
Physics (KL FAMO).

J.K.\ conducted the numerical simulations and analytical calculations with help from M.Ł.\ and W.G.; B.T.\ and J.K.\ performed the stability analysis of solitons. M.Ł.\ conceptualized the research and co-supervised it with K.P.; K.P.\ was responsible for project administration and acquisition of the financial support. All authors took part in the discussion of results. J.K.\ wrote the first draft of the manuscript with help from M.Ł.

\end{acknowledgments}

\appendix
\section{Numerical tools\label{app:A}}
We solve the GGP equation for a complex orbital $\psi(x)$ discretized on a spatial grid with $N_x$ points and spacing $DX=L/N_x$. Parameter $L$ is the box size. We use periodic boundary conditions 
by imposing $\psi(-L/2)=\psi(L/2)$ in every iteration. When we simulate dark solitons inside the droplet we change the phase $\phi(x)$ to $\phi(x)+\frac{2\phi_\infty}{L}x$. The real-time evolution is done with the split-step method. The evolution with the kinetic term is done in the momentum domain, whereas the contact interaction term is calculated in the spatial domain. We do not use any external potential. 
The program \texttt{MUDGE}, written in C++ and implementing the algorithm above, is publicly available (see \hyperlink{sec:data-avail}{Data availability} for link).
The W-DATA format dedicated to store data in
numerical experiments with ultracold Bose and Fermi
gases is used. The W-DATA project is a part of the W-SLDA
toolkit~\cite{WSLDAToolkit}.

\section{Bogoliubov-de Gennes analysis of the solitonic profiles\label{app:B}}

In order to examine the stability of the solitonic solutions as in~\eqref{eq:ansatz}, we linearize the dimensionless generalized Gross-Pitaevskii equation~\eqref{eq:dimless} using the following Ansatz
\begin{equation}
    \Phi(x,t)=\left[\Phi_{0}(x)+\delta\Phi(x,t) \right]e^{-i\mu t},
\end{equation}
where $\Phi_{0}(x)$ is the solitonic wave function, obtained numerically from Eqs.~\eqref{eq:de_rho} and~\eqref{eq:de_phi}. We look for solutions having the form:
\begin{equation}
    \delta\Phi(x,t)=u(x,t)e^{-i\omega t}+v^{\star}(x,t)e^{i\omega t}.
\end{equation}
The linearization of~\eqref{eq:dimless} yields a set of equations, which are formally equivalent to Bogoliubov-de Gennes equations:

\begin{equation} \label{bdg_ham}
\begin{pmatrix}
\hat{h}(x) & \chi(x)  \\
-\chi^\star(x) & -\hat{h}(x)
\end{pmatrix} 
\begin{pmatrix}
u_{n}(x)\\
v_{n}(x)
\end{pmatrix}=
\omega_{n}
\begin{pmatrix}
u_{n}(x)\\
v_{n}(x)
\end{pmatrix}.
\end{equation}

\begin{figure}[h!]
    \centering
    \includegraphics{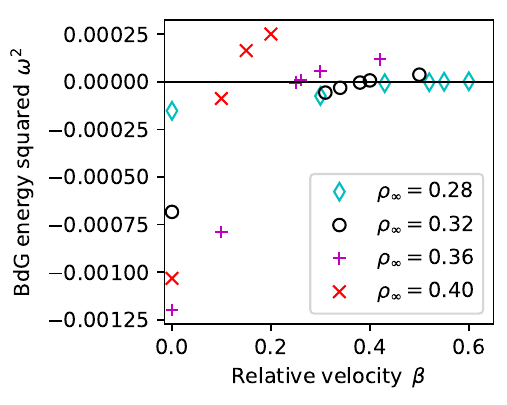}
    \caption{BdG energy squared $\omega^2$ as a function of the soliton relative velocity for different background densities $\rho_\infty$. }
    \label{fig:BdG}
\end{figure}

Here, $\hat{h}(x) = -\partial_x^2/2-\mu+2|\Phi_{0}(x)|^2-\frac{3}{2}|\Phi_{0}(x)|$, and $\chi(x) =  \Phi_{0}^2(x) - \frac{\Phi_{0}^2(x)}{2|\Phi_{0}(x)|}$. Note that the term $\Phi_{0}(x)$ should be treated carefully due to the non-zero imaginary part of the solitonic wave function. The operator $\partial_x^2$ is evaluated using the discrete variable representation~\cite{Bulgac_2013, Jin_2021}.
The BdG Hamiltonian in~\eqref{bdg_ham} is not Hermitian, therefore, it may result in imaginary eigenvalues, which indicate instability. In figure~\ref{fig:BdG}, we examine lowest eigenvalues $\omega$, as a function of soliton velocity for a series of different densities in order to estimate the transition from an unstable profile to a stable one; a negative value of $\omega^2$ indicates the solitonic profile is unstable. It is worth noting the values obtained for $\beta=0$ are consistent with~\cite{Katsimiga_2023b}.

The lack of quantitative agreement in figure~\ref{fig:stability} might be caused by the numerical imperfections, especially related to the numerical grid size. Here, we show results for a numerical grid with 2048 nodes, i.e. resulting in 4096 x 4096 matrix to diagonalize. However, it is noteworthy to mention that the results obtained with grid sizes of 512 and 1024 nodes exhibited comparatively poorer performance. This suggests that the choice of a coarser grid significantly impacted the accuracy of the outcomes, reinforcing the importance of an appropriately refined numerical grid for obtaining more reliable numerical results.
As the grid size increases, so does the dimension of the matrix that needs to be diagonalized. This relationship highlights that larger grids entail more complex computational processes due to the higher-dimensional nature of the underlying matrices. This seems to be the main limitation of our BdG analysis.

\bibliography{sample}

\end{document}